\documentclass{article}

\title{LSHTC: A Benchmark for Large-Scale Text Classification}
\author{Ioannis Partalas, Aris Kosmopoulos, Nicolas Baskiotis\\ Thierry Artieres, George Paliouras, Eric Gaussier\\  Ion Androutsopoulos, Massih-Reza Amini, Patrick Galinari}
\date{}

\usepackage{tikz}
\usetikzlibrary{arrows,positioning,shapes}
\usetikzlibrary{shapes.misc,matrix,fit,backgrounds,calc}
\usepackage{graphicx}
\usepackage{multirow}
\usepackage{amssymb}
\usepackage{subcaption}
\usepackage{url}

 \makeatletter
 \newcommand{\resettranslate}{\let\translate\@firstofone}
 \makeatother

\makeatletter
 \tikzset{circle split part fill/.style  args={#1,#2}{%
  alias=tmp@name, 
   postaction={%
     insert path={
      \pgfextra{%
      \pgfpointdiff{\pgfpointanchor{\pgf@node@name}{center}}%
                   {\pgfpointanchor{\pgf@node@name}{east}}%
      \pgfmathsetmacro\insiderad{\pgf@x}
       \fill[#1] (\pgf@node@name.base) ([xshift=-\pgflinewidth]\pgf@node@name.east) arc
                           (0:180:\insiderad-\pgflinewidth)--cycle;
       \fill[#2] (\pgf@node@name.base) ([xshift=\pgflinewidth]\pgf@node@name.west)  arc
                            (180:360:\insiderad-\pgflinewidth)--cycle;            
          }}}}}
\makeatother

\begin{document}

\maketitle

\begin{abstract}
LSHTC is a series of challenges which aims to assess the performance of classification systems in large-scale classification in a a large number of classes (up to hundreds of thousands).	This paper describes the dataset that have been released along the LSHTC series. The paper details the construction of the datsets and the  design of the tracks as well as the evaluation measures that we implemented and a quick overview of the results. All of these datasets are available online and runs may still be submitted on the online server of the challenges.
\end{abstract}

	\section{Introduction}

Statistical learning has emerged in recent years as a key technology for processing and analyzing large amounts of data. Meanwhile, the growth of such data, their
complexity, and the multiplication of needs generate new data processing problems that cannot be handled within the conventional
frameworks of learning. For example, many applications require
the classification with tens of thousands of classes.

Hierarchical classification is one particular problem of interest of this kind. Indeed hierarchies have become ever more popular for the organization of text documents, particularly on the Web. Web directories and Wikipedia are two examples of such hierarchies. Along with their widespread use, comes the need for automated classification of new documents to the categories in the hierarchy. As the size of the hierarchy grows and the number of documents to be classified increases, a number of interesting machine learning problems arise. In particular, it is one of the rare situations where data sparsity remains an issue, despite the vastness of available data: as more documents become available, more classes are also added to the hierarchy, and there is a very high imbalance between the classes at different levels of the hierarchy. Additionally, the statistical dependence of the classes poses challenges and opportunities for new learning methods. In this specific context the major challenges are: 
\begin{itemize}
 \item 	The development of algorithms capable of scaling to very large number of classes. For example DMOZ is a large web repository, containing over one million categories. 
  \item Taking into account the complex relationships between these categories. For example, the online encyclopedia Wikipedia has more than 20,000 categories related to each other by different types of relationships. 
\end{itemize}

A number of scientific events have been dedicated to this field, including for instance the BioASQ\footnote{http://www.bioasq.org} challenge on large-scale biomedical semantic indexing and question answering, or the international challenges for images (the series of  ImageNet Large Scale Visual Recognition Challenge \footnote {http://www.image-net.org/challenges/LSVRC/2014/}).
Additionally, several workshops treated this problem like 
the Extreme Classification NIPS 2013 workshop \footnote {http://research.microsoft.com/en-us/um/people/manik/events/xc13/} and the WSDM 2014 workshop on Web-Scale Classification: Classifying Big Data from the Web \footnote{http://lshtc.iit.demokritos.gr/WSDM\_WS} 


The LSHTC initiative is a series of challenges on hierarchical text classification which aims to assess the performance of classification systems in large-scale classification in a a large number of classes (up to hundreds of thousands). . It includes tracks of various scales in terms of classes, from thousands to hundreds of thousands as well as many flavors of the large number of classes classification problem, the standard classification problem as well as a multi-task or an unsupervised settings. The many tracks have been designed based on two main corpora from Wikipedia (www.wikipedia.org) and from the ODP Web directory data (www.dmoz.org).	 The LSHTC training datasets may be downloaded on the permanent website \footnote {http://lshtc.iit.demokritos.gr/} where one may still submit a run on the (unavailable) test datasets and gets its performances ranked among the existing participating systems.
		
\section{LSHTC Dataset}
    	
\subsection{Basic Datasets}        

The data used in the LSHTC challenges originates from two popular sources: the DBpedia\footnote {http://dbpedia.org/About} and the ODP (Open Directory Project) directory, also known as DMOZ\footnote {http://www.dmoz.org/}. DBpedia instances were selected from the english, non-regional Extended Abstracts provided by the DBpedia site. The DMOZ instances consist of either \textit{Content} vectors, \textit{Description} vectors or both. A \textit{Content} vectors is obtained by directly indexing the web page using standard indexing chain (pre-processing, stemming/lemmatization, stop-word removal). A \textit{Description} vectors is created by indexing the ODP descriptions of the web pages, which are manually created by the ODP editors.

\subsubsection{Data Creation}

Each dataset is provided in a sparse vector format file, where each line corresponds to an instance. Here is an example of an instance in sparse vector format:

\begin{verbatim} 
5 0:10 8:1 18:2 54:1 442:2 3784:1 5640:1 43501:1
\end{verbatim}
The first numbers (5 in the example) corresponds to the category of the instance. In case of multi-label classification comma-separated numbers are used instead in order to define the categories of and instance.

Each set of numbers separated by `:` correspond to a (feature,value) pair of the vector, where the first number is the feature's id and the second number its frequency (for example feature with the id 18 appears 2 times in the instance). Any feature with the id 0 is not actually a real feature, but is used instead for internal indexing and should by ignored during classification. 

Each token and category of a dataset are mapped to unique numbers. Since each year of the challenge consists of several tracks. In each track a different mapping was used for tokens and categories so that no information could be carried over between tracks. Instances of a track are either split into training, validation and test data or just training and test data. If a validation file is not provided in a track, participants are free to create one, using a subset of the training file. All test instances belong to category 0, meaning that their true label is kept a secret from the participants.

For each track dataset a hierarchy file is also provided. This files contains the parent-child relations between each category of the dataset. During the first hallenge this file was in the form of paths from root to leaf. During the following challenges each line of the file describes a relation between a parent and a child node. 

The hierarchy of the DMOZ datasets is a tree (each node has only one parent) with a maximum depth of 5. All instances deeper than 5 nodes from root are assigned to their ancestor in depth 5. 

The hierarchy of the DBpedia datasets is a graph (a node can have more than one parents), which contains cycles. In some datasets these cycles were removed by ignoring nodes that have been already visited in paths from root to leaf (any parent child relation that would lead to cycle was omitted).
In most datasets classification is only allowed to leaf nodes. In case some inner node of the hierarchy contained any instance, an artificial new node was added as a child of this node and all instances belonging to the initial node were reassigned to the new one.

\subsubsection {LSHTC1}
\label{lshtc1}

The tracks of the first year of the challenge were based on the DMOZ dataset (tree hierarchy) using only single-label instances. The challenge was split into 4 tracks which were composed by different combinations between \textit{Content} and \textit{Description} vectors. Since both types of vectors were used in this challenge only the intersection of the two sets of instances were used for this challenge (we used only instances which had both a \textit{Content} and \textit{Description} vector). In Table \ref{lshtc_pertrack} we present which type of vectors were given in each track for training and test. 

For the tracks of the first challenge a smaller (dry-run) dataset is also provided to facilitate the development of new systems. This dataset is a subpart of its respective main one, but with different mappings between categories and tokens. Another main difference between dry-run and main datasets is that the categories of the dry-run dataset in each sparse vectors are not replaced by 0, since participants are not evaluated in them.

In Table \ref{lshtc_stats_pertrack} we present some statistics regarding the main and dry-run datasets of the fisrt LSHTC. 

In all these datasets classification was only allowed to leaf nodes. 

\begin{table}
\centering
	\begin{tabular}{|l|c|c|c|c|}
		\hline
		\multirow{2}{*} {Track Name} & \multicolumn{2}{|c|}{Content} & \multicolumn{2}{|c|}{Description} \\
		&Train & Test & Train & Test\\
		\hline
		Trak 1: Basic &  \checkmark & \checkmark &  - & -\\
		Trak 2: Cheap & - & \checkmark & \checkmark & -\\
		Trak 3: Expensive &\checkmark&\checkmark&\checkmark&-\\
		Trak 4: Full &\checkmark&\checkmark&\checkmark&\checkmark\\
		\hline	
\end{tabular}
\caption{\textit{Content} and \textit{Description} vectors per track of the first LSHTC.}
\label{lshtc_pertrack}
\end{table}

\begin{table}
\centering
	\begin{tabular}{|l|c|c|}    
    \hline
    Dataset type & Main & Dry-run\\    
    \hline
    Number of categories & 12,294 & 1,139\\
    Number of training instances & 93,805 & 4,463\\
    Number of validation instances & 34.905 & 1.860\\
    Number of test instances & 34.880 & 1,858\\   
    \hline
	\end{tabular}    
    \caption{Statistics regarding each dataset of the first LSHTC.}
\label{lshtc_stats_pertrack}
\end{table}

\subsubsection{LSHTC2}

During LSHTC2, we used multi-label instances and added non-tree hierarchies.
Instead of using, for DMOZ, the intersection between the instances of \textit{Content} and \textit{Description} vector, we decided to keep one of them. We kept the \textit{Content} vectors, since they did not require a human annotator in order be created. Since we decided to move to multi-label classification, we used all the \textit{Content} vectors that we had. 

LSHTC2 consists of three tracks with the first one being the multi-label DMOZ dataset, which as we explained previously has a tree type hierarchy.The other tracks were based on the DBpedia datasets. Track 3 consists of all the extended abstracts, while Track 2 is a subset of the Track 3 dataset with less instances and less categories. While the hierarchy of 
Track 3 contains cycles, the one of Track 2 was cleaned in order be a DAG.

Table \ref{lshtc2_stats_pertrack} presents the main statistics regarding the datasets of LSHTC2.

\begin{table}
\centering
	\begin{tabular}{|l|c|c|c|}    
    \hline
    Dataset & DMOZ & Medium DBpedia & Large DBpedia\\    
    \hline
    Number of categories & 27,875 & 36,504 & 325,056\\
    Number of training instances & 394,754 & 456,886 & 2,365,436\\    
    Number of test instances & 104,263 & 81,262 & 452,167\\   
    Number of stems & 594,158 & 346,299 & 1,617,899\\
    Average categories per inst & 1.02 & 1.86 & 3.26\\
    Deepest leaf in graph & 5 & 10 & 14\\
    \hline
	\end{tabular}    
    \caption{Statistics regarding each dataset of LSHTC2.}
\label{lshtc2_stats_pertrack}
\end{table}

\subsubsection{LSHTC3 and LSHTC4}

The two DBpedia datasets were also used, as Track 1, during the third iteration of the LSHTC challenges (LSHTC3) The only addition was regarding the Medium DBpedia dataset, were we also provided the original text of the instances, without beeing pre-processed. During LSHTC 4, only the Large DBpedia dataset was used for the first track called ``Very Large Supervised Learning'', which was evaluated at Kaggle.\footnote{http://www.kaggle.com/}

            \begin{figure}
            	\centering
                
        \begin{subfigure}[b]{0.45\textwidth}
            	\includegraphics[width=\textwidth]{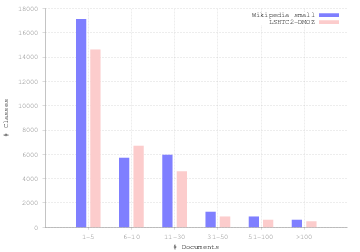}
                \label{fig:class-distr}
                \caption{Class distribution for the DMOZ and Wikipedia small datasets.}      
                \end{subfigure}
                        \begin{subfigure}[b]{0.45\textwidth}
            	\includegraphics[width=\textwidth]{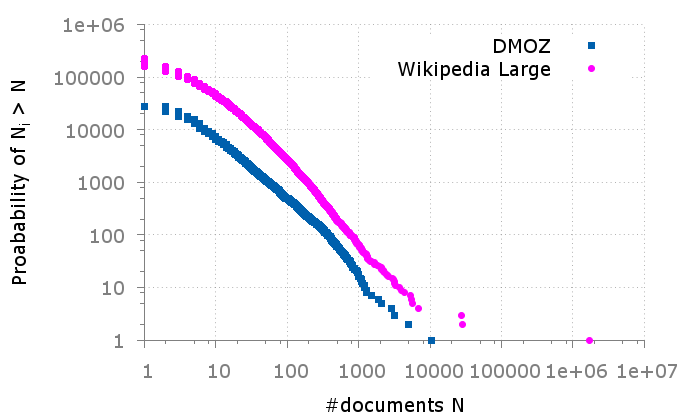}
                \label{fig:class-distr-cum}
                \caption{Class distribution for the DMOZ and Wikipedia small datasets.}      
                \end{subfigure}
                \label{fig:statistics}
                \caption{Statistics on the data sets.}
            \end{figure}

	\section{Evaluation}

		\subsection{Performance Measures}

During the classification tracks of all LSHTC challenges, we used two types of measures in order to evaluate the participating systems, flat and hierarchical. For a flat evaluation measures a prediction can only be correct (if the predicted label is included in the gold set) or wrong (if it is not included in the gold set). On the other hand, hierarchical measures take into account the hierarchical relations of the predicted labels with the gold ones. In that way a prediction can be, right, wrong or partially wrong in various degrees.

Flat evaluation measures are split in to two main families, single-label and multi-label. In single-label classification there is only one gold label for each instance. A system must predict only one label and according to a flat evaluation measure that prediction can be either right or wrong. In order to compute these measures someone must first count the true positives (TP), true negatives (TN), false positives (FP) and false negatives (FN). TP counts how many of the predicted labels were truly gold labels, while TN counts how many of non-predicted labels, were not actually gold. In the same way FP and FN measures the opposite (how many times the respective predictions were wrong).

In single-label classification the most commonly used evaluation measures are accuracy, precision, recall and their $F_1$ measure. 
Accuracy is computed by dividing the correct predictions by the number of instances ($\frac{TP+TN}{TP+TN+FP+FN}$). Precision ($\frac{TP}{TP + FP}$) and recall ($\frac{TP}{TP+FN}$) are computed separately for each category and then averaged over all $M$ categories (Macro versions).  

In multi-label classification, there are several versions of the above flat measures. In the LSHTC challenges we implemented the versions presented in \cite{Tsoumakas_randomk-labelsets}. For example, the micro-F1 measure is calculated as follows:
\[
MiF1 = \frac{2*MiP*MiR}{MiP+MiR},
\]
where $MiP$ and $MiR$ are the micro-precision and micro-recall measures calculated as follows:
\[
MiP = \frac{\sum_{i=1}^{|C|}tp_{c_i}}{\sum_{i=1}^{|C|}(tp_{c_i}+fp_{c_i})}
\]
\[
MiR = \frac{\sum_{i=1}^{|C|}tp_{c_i}}{\sum_{i=1}^{|C|}(tp_{c_i}+fn_{c_i})}
\]
where $tp_{c_i}, fp_{c_i}$ and $fn_{c_i}$ are respectively the true positives, false positives and false negatives for class $c_i$.

Statistical significance tests can be used for all the above flat evaluation measures (S-tests and p-tests) in order to see if there is a statistically significant difference between the performance of two different evaluated systems, according to a respective flat measure. These tests are presented in \cite{Yang99are-examination}.

During the challenges we used and also introduced hierarchical evaluation measures which take into account the relations among the classes. A thorough treatment of hierarchical evaluation measures can be found in \cite{kosmopoulos14}.

		\subsection{The LSHTC Challenge Series}
The LSHTC challenges run from December 2009 until 2014 in four editions, and attracted more than 150 teams from around the world (USA, Europe and Asia).
The results of the challenges were presented in subsequent workshops at the conferences ECIR 2010, ECML 2011, ECML 2012 the challenge being the 
discovery challenge of the conference, WSDM 2014 and ICML 2015. In the following we briefly describe the main technologies that were used in the
competitions. In Table \ref{lshtc_methods} we present the best systems, with some basic information regarding them.

In the first iteration of the LSHTC challenge, there were 4 tracks with 19 participants. All of them participated at the first track, while less than half participated in the other 3 tracks. The most interesting result was that the two best state of the art systems were hierarchical and flat. The first (alpaca, they did not provide a description paper) used hierarchical polynomial SVMs. The second \cite{madani2010large} used online training techniques. Another approach that performed very well was \cite{miao2009hierarchical}, which was using centroids and mildly exploited the hierarchy.

  \begin{table}
\centering
	\begin{tabular}{|l|c|c|c|}    
    \hline
    System & LSHTC & Description & Hierarchical\\    
    \hline
    alpaca & 1 & Polynomial SVMs & \checkmark \\
    \cite{madani2010large} & 1 & Online training & - \\
    \cite{miao2009hierarchical} & 1 & Centroid based & \checkmark \\
	\cite{Brouard11} & 2 & associative network & - \\
    \cite{Wang11} & 2 & Knn with BM25 similarity & - \\
    \cite{Wang12} & 3 & Top-down and meta-features & \checkmark \\
    \cite{Sasaki12} & 3 & Pruning strategy for multi-class & \checkmark \\    
    \cite{Puurula12} & 3 & Ensemble of multinomial naive bayes & -\\
    \cite{Han12} & 3 & k-NN based approach and ranking & - \\
    \cite{Lee12} & 3 & Centroid similarity Rocchio classification & - \\
    \hline
	\end{tabular}    
    \caption{Best systems of LSHTC challenges.}
\label{lshtc_methods}
\end{table}

In LSHTC2, there were 3 tracks with 16 participants. As for LSHTC1, all of them participated at the first track and half participated to the two other tracks. Interestingly, the winning systems in the different tasks are flat : the first one (\cite{Brouard11}), uses an associative network coupled to a post processing of the scores;
the second one (\cite{Wang11}) uses a KNN approach based on a BM25 similarity and a thresholding strategy. Nevertheless, the top tier of systems have very close results, using  hierarchical strategies as well as flat ones.

In LSHTC3, there were 16 participants for the first track and very few for the two other tracks. Flat and hierarchical approaches showed to be competitive. The winning system in the first track is based on the article \cite{Wang12}, a hierarchical approach which has the particularity to consider the multi-class classification as a meta-learning problem based : first a usual top-down hierarchical tree of classifiers is constructed; in the second step, meta-features for a sample are extracted from the scores of each classifier in the tree with the information of the accuracy of the classification as meta-label. Once this meta-learning problem is learned, thresholding strategies are used to classify a sample in the multi-class setting. Another hierachical approach in the top tier is from \cite{Sasaki12} : they consider a usual hierarchical framework, learning a classifier for each edge of the hierarchy, combined to a threshold pruning strategy to improve multi-class classification results. 

The flat classification approaches in the top tier of the results were competitive with the hierarchical ones. \cite{Puurula12} uses a ensemble of multinomial naive bayes with optimization strategies; \cite{Han12} uses a k-NN based approach, by retrieving the most similar training examples and deriving from various scores integrating also the hierarchical information to provide a ranking for each possible class. \cite{Lee12} uses a modification of the Rocchio classification, based on the centroid similarity between an example and the classes, in order to extend the approach to multi-label through label-power set transformation.

	\section{Conclusions}
    This paper presented the LSHTC challenge which run from 2009 to 2014. The goal of the challenge was to asses classification algorithms in a large-scale setting containing hundreds of thousands of target classes. The benchmarks created for the challenges are available for download from the site of LSHTC (\url{http://lshtc.iit.demokritos.gr}) where one can also use the oracles in order to evaluate methods and compare it with the systems that participated in the various editions of the challenge. Our long-term goal is to boost research in large-scale classification by providing a benchmark dataset of reference.

\end{document}